\documentclass[%
reprint,
 superscriptaddress,
showpacs,
 amsmath,amssymb,
 aps,
]{revtex4-1}

\usepackage{graphicx}
\usepackage{dcolumn}
\usepackage{bm}
\usepackage{color}
\usepackage[usenames,dvipsnames]{xcolor}

\usepackage[normalem]{ulem}

\begin{document}

\preprint{APS/123-QED}

\title{A time-dependent Schr\"odinger equation for molecular core-hole dynamics}

\author{A. Pic\'on}
\email{antonio.picon.alvarez@gmail.com}
\affiliation{Argonne National Laboratory, Argonne, Illinois 60439, USA}
\affiliation{Grupo de Investigaci\'on en Aplicaciones del L\'aser y Fot\'onica, Departamento de F\'isica Aplicada, University of Salamanca, E-37008, Salamanca, Spain}
\
 \date{\today}

\begin{abstract}
X-ray spectroscopy is an important tool for the investigation of matter. X rays primarily interact with inner-shell electrons creating core (inner-shell) holes that will decay on the time scale of attoseconds to few femtoseconds through electron relaxations involving the emission of a photon or an electron. The advent of femtosecond x-ray pulses expands x-ray spectroscopy to the time domain and will eventually allow the control of core-hole population on timescales comparable to core-vacancy lifetimes. For both cases, a theoretical approach that accounts for the x-ray interaction while the electron relaxations occur is required. Here we describe a time-dependent framework, based on solving the time-dependent Schr\"odinger equation, that is suitable for describing the induced electron and nuclear dynamics. 
\end{abstract}

\maketitle

\section{Introduction}

For nearly a century, x rays have evolved into an important tool for spectroscopic applications primarily due to their element specificity \cite{Siegbahn1982}. X-ray absorption, emission, and Auger as well as photoelectron spectroscopy  have been used to investigate systems ranging from atoms and molecules in the gas phase \cite{Becker_book} to surfaces, interfaces, and solids \cite{StohrNexafs}.

Over the past decades optical spectroscopy has rapidly progressed towards time-resolved approaches. The advent of femtosecond laser spectroscopy opened the possibility to observe very fast nuclear dynamics and have access to resolve even the vibrational motion of some molecular systems in real time. The wide active area of research that resort to those time-resolved studies is nowadays known as Femtochemistry \cite{Zewail2000}. In the field of Femtochemistry, the common experimental setup is the use of two femtosecond optical lasers, the first one, the pump, excites the molecule while the second one, the probe, probes the induced molecular dynamics. Femtosecond laser spectroscopy has provided real-time studies of dynamics in chemical reactions, materials, and biological systems. 

The field of time-resolved x-ray spectroscopy has developed over the past years \cite{Gessner2015,Chergui2016} and ultra-intense femtosecond pulses from free-electron lasers have opened the door for ultrafast investigations on time-scales similar to core-vacancy decay \cite{Bostedt2016,Bencivenga2015,Yabashi2013,Feldhaus2013}. New approaches for pump/probe techniques involving inner-shell electrons using either optical pump schemes at high-harmonic sources \cite{Chen2014,Fabris2015,Cousin2014,Silva2015,Hergar2016,Leone2016} or xuv \cite{Rudenko2010,Magrakvelidze2012} and x-ray pump schemes at free-electron lasers \cite{Chelsea2015,Ferguson2016,Picon2016,Lehmann2016} are pursued.

From the fundamental point of view, the x-ray interaction of such short pulses with matter yields interesting questions to be explored, as the timescale of the x-ray probing is comparable with the electron relaxation processes triggered by the absorption of the same x-ray pulse. When an x-ray photon is absorbed by a molecule, a core electron is promoted into a highly-excited state leaving behind a core-hole state. Those states are quite unstable and decay rapidly between hundreds of attoseconds to few femtoseconds. Hence, it is possible to tailor the dynamics of the core-hole states before their decay, a unique feature of these ultrashort x-ray pulses. The understanding of this interaction is crucial for the development of unprecedented nonlinear spectroscopy methods with few-femtosecond and attosecond time resolution \cite{Mukamel2009}.

The theoretical models for x-ray spectroscopy are still mostly tailored towards the static case \cite{Becker_book} but in light of the rapidly developing time-resolved x-ray experiments new time-dependent theoretical approaches are needed. In this paper I describe a time-dependent approach that is based on a time-dependent Schr\"odinger equation formalism that includes core-level states, which are relevant to x-ray spectroscopy. Similar time-dependent approaches have been developed in the past in the context of vibrational interference effects on autoionizing electron spectra \cite{Pahl1996}. The approach introduced here can describe both resonant and nonresonant x-ray excitations. The formalism is benchmarked against x-ray absorption and Auger emission data of diatomic molecules and it shows excellent agreement with experimental spectra. With the time-dependent Schr\"odinger approach we have a new tool at hand to describe time-resolved experiments in the x-ray domain that can easily expanded to larger systems.


\section{The theoretical model}
The theoretical approach is based on solving the time-dependent Schr\"odinger equation (TDSE) restricted to those electronic states that are involved in the main dynamics. 
In a molecule, the Hamiltonian may be written in two terms, the electronic and nuclear Hamiltonian
\begin{eqnarray}\label{H_molecule}
\hat{H}_0=\hat{H}_e + \hat{H}_n
\end{eqnarray}
where 
\begin{eqnarray}
\hat{H}_e &=& \sum_j \hat{K}_j +\sum_{ij} \hat{V}_{ij}^{(pe)} + \sum_{j>j'} \hat{V}_{jj'}^{(ee)} \, , \\
\hat{H}_n &=& \sum_i \hat{K}_i + \sum_{i>i'} \hat{V}_{ii'}^{(pp)}\, .
\end{eqnarray}

The wavefunction of the system depends both on the electronic and nuclear coordinates  $\Psi=\Psi({\bf X},{\bf R})$, where ${\bf X}=\{{\bf x_1}, {\bf x_2},... {\bf x_j},...{\bf x_N}\}$ and ${\bf R}=\{{\bf R_1}, {\bf R_2},... {\bf R_i},...{\bf R_{M}}\}$. The wavefunction can be expanded as $\Psi= \sum_{a} b_{a}^{(e)} ({\bf R})\Phi_a^{(e)}({\bf X},{\bf R})$, being $\Phi_a^{(e)}$ an eigenstate of the electronic Hamiltonian for a specific nuclear coordinates ${\bf R}$, that is 
\begin{eqnarray}\label{He_eigensystem}
\hat{H}_e \Phi_a^{(e)}({\bf X},{\bf R}) = E_a^{(e)} ({\bf R}) \Phi_a^{(e)}({\bf X},{\bf R}) \, .
\end{eqnarray}

Calculating the Coulomb electron repulsion of all electrons is an impossible task for molecules having more than two electrons. It is the aim of Quantum Chemistry codes to perform calculations approximating the Coulomb repulsion and obtaining a solution close to considering all electron correlations. In general, we can always assume that the electronic Hamiltonian of the system is composed by two terms
\begin{eqnarray} \label{Heff_Va}
\hat{H}_e = \hat{H}_{\rm eff} + \hat{V}_r
\end{eqnarray}
where $\hat{H}_{\rm eff}$ is the effective Hamiltonian that approximates the electron correlations, and $\hat{V}_r$ is the residual term that is not accounted. As better our approximation to the electron correlations is, as smaller the contribution of the residual potential is. Within the Hilbert space given by the effective Hamiltonian, the wavefunction can be expanded as $\Psi= \sum_{a} b_{a} ({\bf R})\Phi_a({\bf X},{\bf R})$, being $\Phi_a$ now an eigenstate of the effective Hamiltonian
\begin{eqnarray}\label{Heff_eigensystem}
\hat{H}_{\rm eff} \Phi_a({\bf X},{\bf R}) = E_a ({\bf R}) \Phi_a({\bf X},{\bf R}) \, .
\end{eqnarray}

The total Hamiltonian of the molecule is expanded, by using ansatz (\ref{H_molecule}), as
\begin{eqnarray} \label{priorBO}
\hat{H}_0 \Psi = \sum_a \Phi_a({\bf X},{\bf R})  [\hat{H}_n + E_a({\bf R}) + \hat{V}_r({\bf R}) ] b_{a} ({\bf R}) \nonumber \\
+ \sum_{a} b_{a}({\bf R}) [\sum_i \hat{K}_i \Phi_a({\bf X},{\bf R})]  \, .
\end{eqnarray}

In the Born-Oppenheimer (BO) approximation, the change of the nuclear wavepacket is considered much slower in time than the electronic wavepacket, and the second term of Eq. (\ref{priorBO}) is neglected during the time evolution. If we consider the coupling with an electromagnetic wave field $V_I (t)$, the total Hamiltonian $\hat{H}(t)=\hat{H}_0 + V_I (t)$ is time-dependent and then our wavefunction will have an explicit dependence with time as $\Psi (t) = \sum_{a} b_{a} ({\bf R},t)\Phi_a({\bf X},{\bf R})$. Assuming that the external field mainly couples with the electrons, the total Hamiltonian is, within the dipole approximation,
\begin{eqnarray} \label{priorBO_Vt}
&& \hat{H} (t) \Psi (t) =  \nonumber \\
&&\sum_a \Phi_a({\bf X},{\bf R})  [\hat{H}_n + E_a({\bf R}) + \hat{V}_r({\bf R}) + V_I(t) ] b_{a} ({\bf R},t) \nonumber \\
&&+ \sum_{a} b_{a} ({\bf R},t) [\sum_i \hat{K}_i \Phi_a({\bf X},{\bf R})]  . \hspace{.2cm} 
\end{eqnarray}

The time evolution of the quantum system will be described by the time-dependent Schr\"odinger equation $i {\partial \Psi (t)}/{\partial t} = \hat{H} (t) \Psi (t)$, using equation (\ref{priorBO_Vt}) this is
\begin{eqnarray} \label{tdse}
i \sum_{a} {\dot b}_{a} ({\bf R},t)\Phi_a({\bf X},{\bf R}) = \nonumber \\
\sum_a \Phi_a({\bf X},{\bf R})  [\hat{H}_n + V_I(t) + E_a({\bf R}) + \hat{V}_r({\bf R}) ] b_{a} ({\bf R},t) \nonumber \\
+ \sum_{a} b_{a} ({\bf R},t) [\sum_i \hat{K}_i \Phi_a({\bf X},{\bf R})] . \hspace{0.1cm}
\end{eqnarray}

We can interpret the written time-dependent Schr\"odinger equation as nuclear wavepackets propagating along different potential energy surfaces (pes), and those nuclear wavepackets can jump to different electronic pes via the light interaction and nonadiabatic couplings, but also due to the residual potential $\hat{V}_r$ not accounted in the effective Hamiltonian. Note that the nuclear wavepacket amplitudes $b_{a} ({\bf R},t)$ are in the space representation, instead of using the conventional expansion in vibrational states. This has a numerical advantage in solving the TDSE for ultrashort pulses. In the common ultrafast experiments, the molecule is in the ground state or some low-lying excited state. The ultrashort pulse excites the molecule into several states, but due to the localized action of the light-interaction coupling, the excited superposition is well localized in space. If the timescale of the interaction is on the order of hundreds of femtoseconds, we can contain the entire wavefunction in a small spatial grid. With this spatial representation we do not need to calculate explicitly then vibrational or dissociative states.

Solving the complete TDSE is quite demanding, an for numerical purpose, it is a better strategy to limit the electronic states to those that are important during the time evolution of the system. Similar models have been used before, see for example Refs. \cite{Pahl1996,Demekhin_PRA_2011,Lehmann2016}. In the following, we discuss in detail two particular cases of the time-dependent Schr\"odinger equation for inner-shell dynamics (TDSE-IS), the nonresonant and resonant core excitation with Auger decay. However, this approach is quite general and it can be extended to more complex systems by considering more electronic states.

\subsection{Nonresonant core excitation}
In this section we consider the physical scenario that a core electron is ionized, leaving behind a core-hole state. We will assume that the core-hole state mainly decays by Auger processes (this is the case for light atomic elements). We restrict the ansatz of the system to
\begin{eqnarray} \label{ansatz_Auger_1}
\psi (t) = b_0 ({\bf R},t) \Phi_{0} ({\bf X},{\bf R})  + \sum_\varepsilon \sum_i b_{\varepsilon ; i}({\bf R},t) \Phi_{\varepsilon; i} ({\bf X},{\bf R}) + \nonumber\\ 
\sum_{\varepsilon \varepsilon_a} \sum_{i j}  b_{\varepsilon \varepsilon_a ; i j}({\bf R},t) \Phi_{\varepsilon \varepsilon_a ; i j} ({\bf X},{\bf R}) \; , \hspace{0.5cm}
\end{eqnarray}
where $b_0$ stands for the amplitude of the ground state, $b_{\varepsilon ; i}$ for the core-hole states after x-ray photoionization, and $b_{\varepsilon \varepsilon_a ; i j}$ for the final states after Auger decay. Using the ansatz (\ref{ansatz_Auger_1}) in Eq. (\ref{tdse}), and projecting onto an specific electronic state and integrating over the electron coordinates we obtain a system of equations of motion (EOM) for the amplitudes
\begin{eqnarray} 
i\, \dot{b}_0 ({\bf R},t) &=& [\hat{H}_n + E_0({\bf R})]\, b_0 ({\bf R},t) + \nonumber \\
& &\sum_\varepsilon \sum_i  \langle 0 \vert V_I(t) \vert \varepsilon; i \rangle \, b_{\varepsilon ; i}({\bf R},t)  \; , \nonumber \\ \label{EOM_Auger_RPA_1} \\
i\, \dot{b}_{\varepsilon ; i}({\bf R},t) &=& [\hat{H}_n + E_{\varepsilon ; i}({\bf R})] {b}_{\varepsilon ; i}({\bf R},t) + \nonumber \\
&& \langle \varepsilon; i \vert V_I(t) \vert 0 \rangle \, b_0 ({\bf R},t) + \nonumber \\
&& \sum_{\varepsilon' \neq \varepsilon} \sum_{i'\neq i} \langle  \varepsilon; i \vert V_{r}\vert \varepsilon'; i'  \rangle \,  b_{\varepsilon'; i' }({\bf R},t) +  \nonumber \\
&& \sum_{\varepsilon' \varepsilon'_a} \sum_{i' j'} \langle  \varepsilon; i \vert V_{r}\vert  \varepsilon' \varepsilon'_a ; i' j'  \rangle \,  b_{\varepsilon' \varepsilon'_a ; i' j'}({\bf R},t) \; , \nonumber \\ \label{EOM_Auger_RPA_2}\\
i\, \dot{b}_{\varepsilon \varepsilon_a ; i j} ({\bf R},t) &=&  [\hat{H}_n + E_{\varepsilon \varepsilon_a ; i j}({\bf R})] \, b_{\varepsilon \varepsilon_a ; i j} ({\bf R},t) +  \nonumber \\
&& \sum_{\varepsilon'} \sum_{i'} \langle  \varepsilon \varepsilon_a ; i j \vert V_{r}\vert  \varepsilon'; i'  \rangle \,  b_{\varepsilon' ; i'}({\bf R},t) + \nonumber \\
&& \sum_{\varepsilon' \varepsilon'_a \neq \varepsilon \varepsilon_a } \sum_{i' j' \neq i j} \langle  \varepsilon \varepsilon_a ; i j \vert V_{r}\vert  \varepsilon' \varepsilon_a' ; i' j'  \rangle \times  \nonumber \\ 
&& b_{\varepsilon' \varepsilon'_a ; i' j'}({\bf R},t)
\;, \nonumber \\ \label{EOM_Auger_RPA_3}
\end{eqnarray}
where
\begin{eqnarray*}
\langle a \vert V_I(t) \vert a' \rangle = \int d{\bf X}\;\; \Phi_a^*({\bf X},{\bf R}) \; V_I(t) \; \Phi_{a'} ({\bf X},{\bf R})  \\
\langle a \vert V_r \vert a' \rangle = \int d{\bf X}\;\; \Phi_a^*({\bf X},{\bf R}) \; V_r({\bf R}) \; \Phi_{a'} ({\bf X},{\bf R}) 
\end{eqnarray*}

The energies of the ground state, core-excited states, and final states are given by $E_0$, $E_{\varepsilon; i}$, and $E_{\varepsilon \varepsilon_a ; i j}$ respectively. We have neglected the nonadiabatic coupling in Eq. (\ref{tdse}), also the terms 
\begin{eqnarray*}
&&\langle \varepsilon; i \vert  \hat{V}_{r} \vert 0 \rangle \approx 0 \\
&&\langle \varepsilon \varepsilon_a ; i j \vert  \hat{V}_{r} \vert 0 \rangle \approx 0
\end{eqnarray*}
are not considered, as those are quite small compared to the other dominant terms that we discuss in the following.

In core-shell ionization, when the ionization may come from several degenerate states or close by in energies, for example the ionization of 3d electrons in Xe or C 1s electrons in acetylene, {then the Random Phase Approximation (RPA) at the Hartree-Fock level or the multichannel Hartree-Fock theory provide a good theoretical description of the involved electron correlations, see for example Refs. \cite{Becker_book,Yabushita1987,Lucchese1991,Lin2001,Cherepkov2007}. The RPA has also been applied at the level of algebraic-diagrammatic construction (ADC) \cite{Schirmer1996}. The RPA can also be applied in the calculations of Auger decay transitions \cite{Aberg1982}. In these approaches, the coupling between different channels in the final state are considered}. In this work, we consider those electron-correlation couplings to be zero, that is
\begin{eqnarray*}
&& \langle  \varepsilon; i \vert V_{r}\vert \varepsilon'; i'  \rangle \approx 0 \\
&& \langle  \varepsilon \varepsilon_a ; i j \vert V_{r}\vert  \varepsilon' \varepsilon_a' ; i' j'  \rangle  \approx 0
\end{eqnarray*}

The system of equations (\ref{EOM_Auger_RPA_1}), (\ref{EOM_Auger_RPA_2}), and (\ref{EOM_Auger_RPA_3}) can be further decoupled by using the adiabatic approximation, also known as local approximation \cite{Cederbaum1981,Domcke1991}. The adiabatic approximation can be applied to the ionization step, this is known in the Quantum Optics community as Markov approximation \cite{Knight1990,Tannoudji1977}, and to the Auger decay step. Within these approximations, see more details in appendix \ref{sec:appendixA}, the EOMs can be reduced with the derivation of decay rates $\Gamma$ that accounts for the ionization of the ground state and the Auger decay of the core-hole state, and Stark shifts $R$ that account for the dephasing introduced by the continuum part that has been decoupled:
\begin{widetext}
\begin{eqnarray}  \nonumber
&& i\, \dot{b}_0 ({\bf R},t) = [\hat{H}_n + E_0({\bf R})]\, b_0 ({\bf R},t)  -i \left[ {\Gamma_I({\bf R},t) \over 2} + i R_I ({\bf R},t) \right] b_0 ({\bf R},t)  \\
&& - \sum_\varepsilon \sum_i  \langle 0 \vert V_I \vert \varepsilon;i \rangle \int_{t_0}^t dt'  \sum_{\varepsilon'\neq\varepsilon} \sum_{i'\neq i} \left[ {\Gamma_{\varepsilon i,\varepsilon' i'}({\bf R}) \over 2} + i R_{\varepsilon i,\varepsilon' i'}({\bf R}) \right] b_{\varepsilon' ; i'}({\bf R},t')   \; e^{-i (E_{\varepsilon ; i}({\bf R}) + R_{\varepsilon i, \varepsilon i}({\bf R}) - i \Gamma_{\varepsilon i, \varepsilon i}({\bf R})/2) (t-t')}   \; , \nonumber \\
&& i\, \dot{b}_{\varepsilon ; i}({\bf R},t)  = [\hat{H}_n + E_{\varepsilon ; i}({\bf R})] {b}_{\varepsilon ; i}({\bf R},t) + \langle \varepsilon; i \vert V_I(t) \vert 0 \rangle \, b_0 ({\bf R},t)   
-i \sum_{\varepsilon'} \sum_{i'} \left[ {\Gamma_{\varepsilon i,\varepsilon' i'}({\bf R}) \over 2} + i R_{\varepsilon i,\varepsilon' i'}({\bf R}) \right] b_{\varepsilon' ; i'}({\bf R},t)  \; , \nonumber \\
&& i\, \dot{b}_{\varepsilon \varepsilon_a ; i j} ({\bf R},t) =  [\hat{H}_n + E_{\varepsilon \varepsilon_a ; i j}({\bf R})] \, b_{\varepsilon \varepsilon_a ; i j} ({\bf R},t) +  
\sum_{\varepsilon'} \sum_{i'} \langle  \varepsilon \varepsilon_a ; i j \vert V_{r}\vert  \varepsilon'; i'  \rangle \,  b_{\varepsilon' ; i'}({\bf R},t) 
\;, \nonumber \\
\label{EOM_Auger_adb}
\end{eqnarray}
where
\begin{eqnarray*}
{\Gamma_{\varepsilon i,\varepsilon' i'} ({\bf R}) \over 2} &=& \pi \sum_{\varepsilon'' \varepsilon''_a} \sum_{i'' j''} \langle  \varepsilon; i \vert V_{r}\vert  \varepsilon'' \varepsilon''_a ; i'' j''  \rangle \, 
\langle  \varepsilon'' \varepsilon''_a ; i'' j'' \vert V_{r}\vert  \varepsilon'; i'  \rangle  \;\; \delta(E_{\varepsilon'' \varepsilon''_a ; i'' j''} - E_{\varepsilon' ; i'}) \\
R_{\varepsilon i,\varepsilon' i'}({\bf R}) & = & -i \sum_{\varepsilon'' \varepsilon''_a} \sum_{i'' j''} \langle  \varepsilon; i \vert V_{r}\vert  \varepsilon'' \varepsilon''_a ; i'' j''  \rangle \, 
\langle  \varepsilon'' \varepsilon''_a ; i'' j'' \vert V_{r}\vert  \varepsilon'; i'  \rangle \;\; P \left[ {1 \over E_{\varepsilon'' \varepsilon''_a ; i'' j''} - E_{\varepsilon' ; i'}}\right] \\
{\Gamma_{I}({\bf R},t) \over 2} & = & {\Omega^2(t)\over4} \sum_\varepsilon \sum_i  \langle 0 \vert \tilde{V}_I \vert \varepsilon; i \rangle  \langle \varepsilon; i \vert \tilde{V}_I \vert 0 \rangle 
{\Gamma_{\varepsilon i,\varepsilon i}/ 2 \over (E_{\varepsilon; i} +R_{\varepsilon i,\varepsilon i} -E_0 -\omega)^2 + \left({\Gamma_{\varepsilon i,\varepsilon i}/ 2}\right)^2 } \, \\
R_{I}({\bf R},t)  & = & - {\Omega^2(t)\over4} \sum_\varepsilon \sum_i  \langle 0 \vert \tilde{V}_I \vert \varepsilon; i \rangle  \langle \varepsilon; i \vert \tilde{V}_I \vert 0 \rangle 
{E_{\varepsilon; i} +R_{\varepsilon i,\varepsilon i} -E_0 -\omega \over (E_{\varepsilon; i} +R_{\varepsilon i,\varepsilon i} -E_0 -\omega)^2 + \left({\Gamma_{\varepsilon i,\varepsilon i}/ 2}\right)^2 }
\end{eqnarray*}
\end{widetext}
The symbol $P$ stands for the principal value. The Rabi frequency of the pulse is given by $\Omega(t)$, the frequency of the pulse by $\omega$, the dipole moments between ground state and core-hole states by $\langle 0 \vert \tilde{V}_I \vert \varepsilon; i \rangle$, where $\tilde{V}_I$ stands for the electric dipole moment, i.e. $- \sum_j q_j\; {\bf r}_j \cdot {\bf s}$ where ${\bf s}$ is the polarization direction. The ionization rate of the ground state is related to the term $\Gamma_{I}(t)$, which depends on the envelope (intensity) of the pulse. The Auger transitions are given by the couplings $\langle  \varepsilon; i \vert V_{r}\vert  \varepsilon'' \varepsilon''_a ; i'' j''  \rangle$. The decay of the core-excited state is related to the term $\Gamma_{\varepsilon i,\varepsilon i}$, which is the sum of all Auger transitions allowed in the system. 

\subsection{Resonant core excitation}
In this section we consider the physical scenario in which a core electron is promoted into a bound highly-excited state, leaving behind a core-hole state. After core resonant excitation, the system is still neutral (no loss of electrons). We will assume that the core-hole state mainly decays by Auger processes. Similarly to the nonresonant case, we limit the Hilbert space to the electronic states mainly involves in the dynamics. We consider the ansatz of the system to be
\begin{eqnarray} \label{ansatz_Auger_2}
\psi (t) = b_0 ({\bf R},t) \Phi_{0} ({\bf X},{\bf R})  + b_{c}({\bf R},t) \Phi_{c} ({\bf X},{\bf R}) + \nonumber\\ 
\sum_{\varepsilon_a} \sum_{i j}  b_{c \varepsilon_a ; i j}({\bf R},t) \Phi_{c \varepsilon_a ; i j} ({\bf X},{\bf R}) \; , \hspace{0.5cm}
\end{eqnarray}
where $b_0$ stands for the amplitude of the ground state, $b_{c}$ for the core-hole state after resonant excitation, and $b_{c \varepsilon_a ; i j}$ for the final states after Auger decay. Using the ansatz (\ref{ansatz_Auger_2}) in Eq. (\ref{tdse}), projecting onto an specific electronic state and integrating over the electron coordinates, we obtain the EOM for the resonant excitation
\begin{eqnarray} 
i\, \dot{b}_0 ({\bf R},t) &=& [\hat{H}_n + E_0({\bf R})]\, b_0 ({\bf R},t) \nonumber \\
&& +  \langle 0 \vert V_I(t) \vert c \rangle \, b_{c}({\bf R},t)  \; , \nonumber \\
i\, \dot{b}_{c}({\bf R},t) &=& [\hat{H}_n + E_{c}({\bf R})] {b}_{c}({\bf R},t) \nonumber \\
&&+  \langle c \vert V_I(t) \vert 0 \rangle \, b_0 ({\bf R},t) \nonumber \\
&&+ \sum_{\varepsilon'_a} \sum_{i' j'} \langle c \vert V_{r}\vert  c\, \varepsilon'_a ; i' j'  \rangle \,  b_{c \varepsilon'_a ; i' j'}({\bf R},t) \; , \nonumber \\ 
i\, \dot{b}_{c \varepsilon_a ; i j} ({\bf R},t) &=&  [\hat{H}_n + E_{c \varepsilon_a ; i j}({\bf R})] \, b_{c \varepsilon_a ; i j} ({\bf R},t)  \nonumber \\
&& + \, \langle  c\, \varepsilon_a ; i j \vert V_{r}\vert c \rangle \,  b_{c}({\bf R},t)  \nonumber \\ 
&& + \sum_{\varepsilon'_a \neq  \varepsilon_a } \sum_{i' j' \neq i j} \langle  c \, \varepsilon_a ; i j \vert V_{r}\vert  c \, \varepsilon_a' ; i' j'  \rangle \times  \nonumber \\ 
&& b_{c \varepsilon'_a ; i' j'}({\bf R},t)
\label{EOM_Auger_adb_R}
\end{eqnarray}
We can decouple the EOMs by using the adiabatic approximation in the Auger step and further reduce the EOM by using decay rates $\Gamma$ and Stark shifts $R$ parameters. Within the adiabatic approximation, neglecting the RPA terms, we obtain
\begin{eqnarray} 
i\, \dot{b}_0 ({\bf R},t) &=& [\hat{H}_n + E_0({\bf R})]\, b_0 ({\bf R},t) \nonumber \\
&& +  \langle 0 \vert V_I(t) \vert c \rangle \, b_{c}({\bf R},t)  \; , \nonumber \\ 
i\, \dot{b}_{c}({\bf R},t) &=& [\hat{H}_n + E_{c}({\bf R})] {b}_{c}({\bf R},t) \nonumber \\
&& +  \langle c \vert V_I(t) \vert 0 \rangle \, b_0 ({\bf R},t) \nonumber \\
&&-i  \left[ {\Gamma_c({\bf R}) \over 2} + i R_c ({\bf R}) \right]  {b}_{c}({\bf R},t) \; , \nonumber \\
i\, \dot{b}_{c \varepsilon_a ; i j} ({\bf R},t) &=&  [\hat{H}_n + E_{c \varepsilon_a ; i j}({\bf R})] \, b_{c \varepsilon_a ; i j} ({\bf R},t)  \nonumber \\
&& + \, \langle  c \varepsilon_a ; i j \vert V_{r}\vert c \rangle \,  b_{c}({\bf R},t) \nonumber  \; . \nonumber \\ 
\label{EOM_Auger_adb_R}
\end{eqnarray}
where
\begin{eqnarray*}
{\Gamma_{c} ({\bf R}) \over 2} &=& \pi \sum_{\varepsilon'_a} \sum_{i' j'} \vert \langle c \, \varepsilon'_a ; i' j'  \vert V_{r}\vert  c  \rangle \vert^2 \, 
\delta(E_{c \varepsilon'_a ; i' j'} - E_{c}) \\
R_{c}({\bf R}) & = & -i \sum_{\varepsilon'_a} \sum_{i' j'} \vert \langle  c \, \varepsilon'_a ; i' j' \vert V_{r}\vert c \rangle \vert^2 \; P \left[ {1 \over E_{c \varepsilon'_a ; i' j'} - E_{c}}\right] 
\end{eqnarray*}

\section{Numerical implementation}
In the introduced theoretical model, the time evolution of the system is governed by EOMs such as Eqs. (\ref{EOM_Auger_adb}) and (\ref{EOM_Auger_adb_R}). By calculating all the electronic properties at different nuclear geometries -energies, electric dipole transitions, and Auger dipole transitions-, the numerical problem reduces to solving a system of coupled ordinary differential equations. For that purpose, we can use common numerical methods such as Runge-Kutta or Crank-Nicolson methods.

The electronic calculations can be calculated with standard Quantum Chemistry codes, besides those matrix elements involving continuum orbitals. Most common Quantum Chemistry codes are based on multi-center grids expanded with localized basis set, most commonly Gaussian basis. For calculating the energies of the ground state, core-hole states, and final state, we need first to choose a level of description for the electronic correlations, for example Hartree-Fock (HF), Configuration Interaction (CI), Coupled Cluster (CC), or Multi-Reference CI (MRCI), which determines the Hamiltonian $H_{\rm eff}$ and the residual $V_{r}$. Often, for a better description of the system, we need to use a different level of electron correlations for different electronic states. For example, the core-hole state energies have a high-degree of electron relaxation and a second self-consistent field (SCF) calculation by imposing a hole in the corresponding core orbital results in a much better accuracy \cite{Agren1994,Shirai2004,Besley2009}. If we consider electronic states with different $H_{\rm eff}$, then we need to modify correspondingly the EOMs given by Eqs. (\ref{EOM_Auger_adb}) and (\ref{EOM_Auger_adb_R}) by including terms with overlapping factors, as the eigenstates would not be orthogonal anymore. 

Most Quantum Chemistry codes do not include the possibility to calculate continuum orbitals, which are needed for obtaining matrix elements such as electric dipole (ionization) and Auger transitions. There are several approaches to calculate continuum orbitals, such as Dyson orbital methods \cite{Ortiz1999,Gozem2015} or single-center expansions based on scattering theory \cite{Lucchese1982,Demekhin2011}. A Stieltjes imaging is often use in the literature to obtain observables such as photoionization cross sections \cite{Langhoff1973,Cukras2013} or Auger decay transitions \cite{Averbukh2005}. However, this method do not allow to obtain the continuum waves required for the TDSE-IS. 

The initial state $b_0 ({\bf R},t_0)$ has to be calculated prior to solve the TDSE-IS. Once we have the pes for the ground state, we can diagonalize the nuclear Hamiltonian, in the absence of any external field, to obtain the vibrational states of the ground state, or we can also use an imaginary time-evolution method for this pourpose.

\section{Comparison with experimental data}

By solving the TDSE-IS we can calculate the most relevant observables to be measured in experiments, even in static experiments, such as x-ray absorption or Auger spectra. In the following section we explain in details how to calculate those observables within this time-dependent framework and compare it to previously published experimental results.

\subsection{X-ray absorption spectroscopy}
X-ray absorption spectroscopy is a very common technique at synchrotrons. Nowadays those techniques have been highly refined, mainly due to the experimental advances in selecting and tuning the photon energy of the x-ray pulses with a narrow bandwidth. In a x-ray absorption spectrum we can distinguish two domains: the x-ray absorption near-edge structure (XANES) and the extended x-ray absorption fine structure (EXAFS), corresponding to low and high photon energies respectively. XANES contains information about the resonant excitations and continuum excitations near resonances, providing information about the electronic configuration and local chemical environment with respect to the absorber. EXAFS is the high-energy domain where continuum photoelectrons are dominated by single scattering events, providing information about the coordination number, type, and distance of ligating atoms with respect to the absorber. X-ray absorption spectroscopy is a powerful spectroscopic technique that is used in a wide range of applications ranging from photochemistry and solar energy conversion \cite{Cobalt2016,Nickel2016,Bressler2010,Yano2009}, interfacial electron transfer in photocatalysis and biological enzymatic systems \cite{Zhang2011}, to materials characterization.

\begin{figure}[t]
\begin{center}
\includegraphics[width=7.cm]{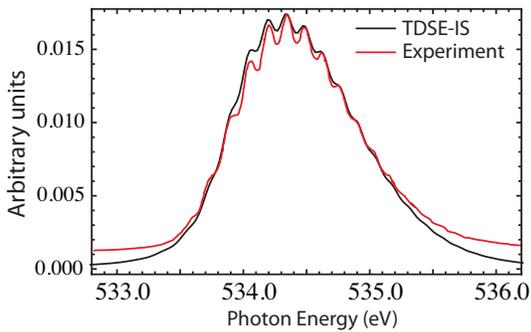}
\caption{X-ray absorption spectrum for CO in the energy range of the O 1s $\rightarrow \pi^*$ resonance. The TDSE-IS was solved for 50-fs x-ray pulses at different photon energies with $10^{13}$ W/cm$^2$ peak intensity. The vibrational states of the electronic 1s$^{-1} \pi^*$ level is resolved. The experimental data is taken from Ref. \cite{Puttner1999}.}
\label{fig:XAS}
\end{center}
\end{figure}

We start discussing the case of static x-ray absorption spectroscopy. The time evolution of the system is mainly given by the nuclear wavepackets amplitudes obtained by solving the TDSE-IS. At the end of the x-ray pulse interaction, the system keeps evolving (electron relaxations and nuclear propagation), but the population in the ground state will remain constant. The difference of population of what we have at the beginning and after the x-ray pulse in the ground state will be related to the absorption signal for a specific photon energy. If we perform the TDSE-IS using x-ray pulses with different photon energies, we can then obtain the x-ray absorption spectrum. We show the calculated x-ray absorption spectrum for carbon monoxide in Fig. \ref{fig:XAS} in the energy range of the O 1s $\rightarrow \pi^*$ resonance. The pes were calculated using the quadruple-zeta Dunning basis cc-pVQZ \cite{Dunning1989} at the level of multi reference configuration interaction (MRCI) by using the quantum chemistry code COLUMBUS \cite{Lischka2001}. In the equilibrium distance the molecule is well-described by a single reference, and it is a good approximation to calculate the electric dipole and Auger transitions at the single reference level, see for example Ref. \cite{Piancastelli1997}. We consider a spatial grid for the internuclear distance from 1.2 to 25 a.u., with a spatial resolution of dR = 0.01 a.u.. We solve the TDSE-IS using a fourth-order Runge-Kutta method. We observe that the vibrational structure of the resonance is perfectly resolved in spite of the spatial coordinate representation of the TDSE-IS. The energy spacing matches very well with the experiment, this is mainly due to the good description of the pes for the core-hole state. In the calculations we approximate the electric dipole moment calculated at the equilibrium distance to be equal at all nuclear geometries. As the initial wavepacket in the ground state is well localized, this approximation is quite good and results in a good agreement between the relative peaks of the vibrational states in the spectrum. Note that no detector or natural width broadening have been used in the calculated spectrum, the represented black line is directly obtained from the TDSE-IS calculations. 

Similarly, for time-resolved studies we can keep track of the population in the transient states induced by the pump pulse and then obtain the transient x-ray absorption spectrum by taking the population difference before and after the probe pulse. 

\subsection{Auger electron spectroscopy}
Auger electron spectroscopy is a common technique used in gas-phase experiments and surfaces of condensed matter systems \cite{GrantBook}. This technique is based on detecting the Auger electron emitted after core-hole decay. Because we may select a particular electronic state by detecting the Auger electron, we can retrieve information about the electronic configuration of the system. Also, the Auger electron may be emitted from the valence shell and provides thus information about the local chemical environment with respect to the absorber. 


The calculation of the Auger electron spectrum using the TDSE-IS will be slightly different for the resonant and nonresonant core excitation. We start discussing the nonresonant case, in which two electrons are located in the continuum; the photoelectron and the Auger electron. Within the TDSE-IS framework we can calculate the two-electron coincidence measurements, i.e. the measurement of the photoelectron and Auger electron in coincidence, given by
\begin{eqnarray}
P(\varepsilon,\varepsilon_a) = \lim_{t\rightarrow \infty} \sum_{ij} \int\! d{\bf R} \, \vert b_{{\varepsilon}{\varepsilon_a};ij} ({\bf R},t) \vert^2
\end{eqnarray}

In the previous formula, although it is not written explicitly, we consider also the sum over the other quantum numbers of the photoelectron and Auger electron. If we are also interested in the angular resolution of the emitted electrons, then we need to remove the sum over the orbital angular momenta of the continuum orbitals. The photoelectron spectrum and the Auger spectrum are then given by
\begin{eqnarray}
P_{ph}(\varepsilon)=\sum_{\varepsilon_a} P(\varepsilon,\varepsilon_a) \; \label{Ph_Spectrum}, \\
P_{a}(\varepsilon_a)=\sum_{\varepsilon} P(\varepsilon,\varepsilon_a) \; \label{Auger_Spectrum},
\end{eqnarray}
respectively. 
In the resonant case, the previous formulas of the Auger electron spectrum are reduced to 
\begin{eqnarray}
P_a(\varepsilon_a) = \lim_{t\rightarrow \infty} \sum_{ij} \int\! d{\bf R} \, \vert b_{{c}{\varepsilon_a};ij} ({\bf R},t) \vert^2 \, .
\end{eqnarray}

\begin{figure}[t]
\begin{center}
\includegraphics[width=7.cm]{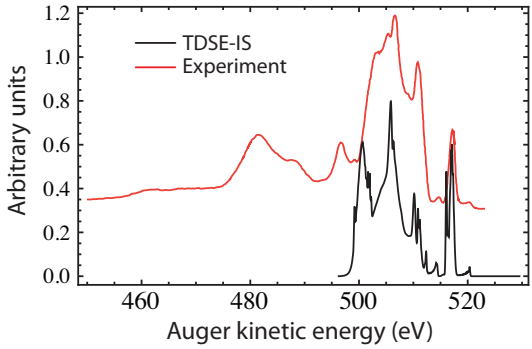}
\caption{Auger electron spectrum for CO excited at 534.5 eV in the O 1s $\rightarrow \pi^*$ resonance. The TDSE-IS was solved for a 50-fs x-ray pulse with $10^{13}$ W/cm$^2$ peak intensity. The experimental data is taken from Ref. \cite{Piancastelli1997}.}
\label{fig:Auger}
\end{center}
\end{figure}

In Fig. \ref{fig:Auger} we show the Auger decay spectrum for CO at 534.5 eV in the O 1s $\rightarrow \pi^*$ resonance. In the TDSE-IS, we have only included the thirteen dominant low-lying-excited states after Auger decay, corresponding to the emission of Auger electrons with high kinetic energies. The calculated Auger spectrum is in good agreement with the experimental spectrum of Ref. \cite{Piancastelli1997}. In the calculated spectrum we are able to observe vibrational structure, while they are smoothed out in the experimental spectrum. This could be due to the 350 meV electron energy resolution of the experiment. Also, the peaks located in the calculated spectrum around 500 eV should be shifted to higher kinetic energy by 3 eV. However, the spectrum is overall well-described and this clearly shows the versatility of the TDSE-IS to obtain observables that can be measured in experiments.

\section{Conclusions}
In conclusion, we have derived a theoretical approach that accounts both for x-ray excitation and electron relaxation of the core hole in a time-dependent framework. This approach allows us to describe and explore the underlying mechanism of few-femtosecond and attosecond x-ray pulses interacting with molecules. This might open the possibility to explore the role of Auger processes in the coherent evolution of the nuclear wavepackets as well as nonadiabatic effects during x-ray excitation. Also, the introduced theoretical approach is ideal for calculating momenta distribution retrieved from electron-ion coincidence measurements, which are very sensitive to both electronic configurations and nuclear geometries. These techniques will be significantly enhanced at future x-ray sources with high-repetition rate capabilities. {The introduced framework can also be extended in order to include the interaction of a strong-field laser with the system and thus study interesting topics of charge migration in molecules with high-harmonic generation \cite{Lepine2014,Leone2016b,Kraus2015}. The strong-field laser is well described within the Strong-Field Approximation (SFA), and the connection of the SFA with a quantum formalism has been shown in Ref. \cite{lewen94A}, which can be adapted to the present framework. This will enable to explore the SFA in molecules, both in the ionization as well as the Auger-decay step \cite{APA_2013_PRA,APA_2013_NJP}. Within this formalism we could also explore electron dynamics induced by the coupling between different core-hole states, analogously to previous approaches used in atomic systems \cite{Pabst2011,Pabst2013}.\\ }


\begin{acknowledgments}
A.P. is grateful to Christoph Bostedt for inspiring this project, scientific discussions, and his support. This material is based upon work supported by the U.S. Department of Energy, Office of Science, Basic Energy Sciences, Chemical Sciences, Geosciences, and Biosciences Division and supported the Argonne group under contract no. DE-AC02-06CH11357. A.P. also acknowledges support from the European UnionÕs Horizon 2020 research and innovation programme under the Marie Sklodowska-Curie grant agreement No 702565.
\end{acknowledgments}

\appendix

\section{Adiabatic approximation} \label{sec:appendixA}

In this section we derive the adiabatic approximation for the Auger transition step. We start taking the integral form of Eq. (\ref{EOM_Auger_RPA_3})
\begin{eqnarray*}
&&b_{\varepsilon \varepsilon_a ; i j} ({\bf R},t) = -i e^{-i[\hat{H}_n + E_{\varepsilon \varepsilon_a ; i j}({\bf R})] t} \times \\
&& \int_{0}^{t} dt' e^{i[\hat{H}_n + E_{\varepsilon \varepsilon_a ; i j}({\bf R})] t'}   \sum_{\varepsilon' i'}  \langle  \varepsilon \varepsilon_a ; i j \vert V_{r}\vert  \varepsilon'; i'  \rangle \,  b_{\varepsilon' ; i'}({\bf R},t) 
\end{eqnarray*}
and including this form into Eq. (\ref{EOM_Auger_RPA_2}) to obtain a new equation without the amplitudes of the final states
\begin{widetext}
\begin{eqnarray*}
&&i\, \dot{b}_{\varepsilon ; i}({\bf R},t) = [\hat{H}_n + E_{\varepsilon ; i}({\bf R})] {b}_{\varepsilon ; i}({\bf R},t) + \langle \varepsilon; i \vert V_I(t) \vert 0 \rangle \, b_0 ({\bf R},t) + \nonumber \\
&&-i \sum_{\varepsilon'' \varepsilon''_a} \sum_{i'' j''} \langle  \varepsilon; i \vert V_{r}\vert  \varepsilon'' \varepsilon''_a ; i'' j''  \rangle \,  
  e^{-i[\hat{H}_n + E_{\varepsilon'' \varepsilon''_a ; i'' j''}({\bf R})] t} \int_{0}^{t} dt' e^{i[\hat{H}_n + E_{\varepsilon'' \varepsilon''_a ; i'' j''}({\bf R})] t'}   \sum_{\varepsilon' i'}  \langle  \varepsilon'' \varepsilon''_a ; i'' j'' \vert V_{r}\vert  \varepsilon'; i'  \rangle \,  b_{\varepsilon' ; i'}({\bf R},t') .
\end{eqnarray*}
\end{widetext}
The second line can be reduced to a decay rate $\Gamma$ factor that accounts for the Auger decay yield and a Stark shift $R$ factor that accounts for the dephasing. In order to perform the integration in the second line, the core-hole amplitude $ b_{\varepsilon' ; i'}({\bf R},t')$ needs to be expressed in the eigenbasis of the operator $[\hat{H}_n + E_{\varepsilon'' \varepsilon''_a ; i'' j''}({\bf R})]$. Therefore, we express the core-hole amplitude as
\begin{eqnarray*}
b_{\varepsilon' ; i'}({\bf R},t') = \sum_\nu c_{\nu,\varepsilon' i'} (t') \,b_{\nu, \varepsilon' i'} ({\bf R}) \\
= \sum_{\nu\nu'} c_{\nu,\varepsilon' i'} (t') \, t^{\nu', \varepsilon'' \varepsilon''_a i'' j''}_{\nu, \varepsilon' i'} \,b_{\nu',\varepsilon'' \varepsilon''_a i'' j''}({\bf R})
\end{eqnarray*}
where first we expand the core-hole state nuclear wavepacket in the core-hole vibrational basis, then every vibrational core-hole state wavefunction is expanded in vibrational states of the electronic level $\varepsilon'' \varepsilon''_a i'' j''$, $t^{\nu', \varepsilon'' \varepsilon''_a i'' j''}_{\nu, \varepsilon' i'}$ being the coefficients of the transformation (related to the Frank-Condon factors). Now we use this expansion in order to convert the operators $[\hat{H}_n + E_{\varepsilon'' \varepsilon''_a ; i'' j''}({\bf R})]$ into energies and be able to perform the integration over time $t'$
\begin{widetext}
\begin{eqnarray*}
&&i\, \dot{b}_{\varepsilon ; i}({\bf R},t) = [\hat{H}_n + E_{\varepsilon ; i}({\bf R})] {b}_{\varepsilon ; i}({\bf R},t) + \langle \varepsilon; i \vert V_I(t) \vert 0 \rangle \, b_0 ({\bf R},t) + \nonumber \\
&&-i \sum_{\varepsilon'' \varepsilon''_a} \sum_{i'' j''} \sum_{\nu\nu'} \langle  \varepsilon; i \vert V_{r}\vert  \varepsilon'' \varepsilon''_a ; i'' j''  \rangle \,  
  e^{-i E_{\nu',\varepsilon'' \varepsilon''_a i'' j''} t} \times \\
&& \int_{0}^{t} dt' t^{\nu', \varepsilon'' \varepsilon''_a i'' j''}_{\nu, \varepsilon' i'} e^{i E_{\nu',\varepsilon'' \varepsilon''_a i'' j''} t'}   \sum_{\varepsilon' i'}  \langle  \varepsilon'' \varepsilon''_a ; i'' j'' \vert V_{r}\vert  \varepsilon'; i'  \rangle \,  c_{\nu,\varepsilon' i'} (t') \,b_{\nu',\varepsilon'' \varepsilon''_a i'' j''}({\bf R}) .
\end{eqnarray*}
\end{widetext}
assuming that $ \langle  \varepsilon'' \varepsilon''_a ; i'' j'' \vert V_{r}\vert  \varepsilon'; i'  \rangle$ slowly changes with ${\bf R}$. Within the adiabatic (Markov) approximation, we split the time-dependent factors in slow and fast time variant:
\begin{eqnarray*}
c_{\nu,\varepsilon' i'}(t') = \tilde{c}_{\nu,\varepsilon' i'}(t') e^{-iE_{\nu,\varepsilon' i'}t'} 
\end{eqnarray*}
and we will have a new integral that can be written as
\begin{eqnarray*}
&&\int_{t_0}^t  dt'  e^{iE_{\nu',\varepsilon'' \varepsilon''_a i'' j''}t'} e^{-iE_{\nu,\varepsilon' i'}t'} \approx \\
&&\pi \, \delta(E_{\nu',\varepsilon'' \varepsilon''_a i'' j''}-E_{\nu,\varepsilon' i'}) \, e^{i(E_{\nu',\varepsilon'' \varepsilon''_a i'' j''}-E_{\nu,\varepsilon' i'})t} \\
&&-i \, P \left[ {1 \over E_{\nu',\varepsilon'' \varepsilon''_a i'' j''}-E_{\nu,\varepsilon' i'}} \right]e^{i(E_{\nu',\varepsilon'' \varepsilon''_a i'' j''}-E_{\nu,\varepsilon' i'})t} \, .
\end{eqnarray*}
where $P$ stands for the principal part. Hence, the integration is split into two terms, and the previous EOM is then reduced to
\begin{widetext}
\begin{eqnarray*}
&&i\, \dot{b}_{\varepsilon ; i}({\bf R},t) = [\hat{H}_n + E_{\varepsilon ; i}({\bf R})] {b}_{\varepsilon ; i}({\bf R},t) + \langle \varepsilon; i \vert V_I(t) \vert 0 \rangle \, b_0 ({\bf R},t) + \nonumber \\
&&-i  \sum_{\varepsilon' i'}  \sum_{\varepsilon'' \varepsilon''_a} \sum_{i'' j''} \sum_{\nu\nu'} \pi \delta(E_{\nu',\varepsilon'' \varepsilon''_a i'' j''}-E_{\nu,\varepsilon' i'}) \, \langle  \varepsilon; i \vert V_{r}\vert  \varepsilon'' \varepsilon''_a ; i'' j''  \rangle \,  
   \langle  \varepsilon'' \varepsilon''_a ; i'' j'' \vert V_{r}\vert  \varepsilon'; i'  \rangle
   c_{\nu,\varepsilon' i'} (t)\,  t^{\nu', \varepsilon'' \varepsilon''_a i'' j''}_{\nu, \varepsilon' i'}  \,b_{\nu',\varepsilon'' \varepsilon''_a i'' j''}({\bf R}) \\
&&- \sum_{\varepsilon' i'}  \sum_{\varepsilon'' \varepsilon''_a} \sum_{i'' j''} \sum_{\nu\nu'} P \left[ {1 \over E_{\nu',\varepsilon'' \varepsilon''_a i'' j''}-E_{\nu,\varepsilon' i'}} \right] \langle  \varepsilon; i \vert V_{r}\vert  \varepsilon'' \varepsilon''_a ; i'' j''  \rangle \,  
   \langle  \varepsilon'' \varepsilon''_a ; i'' j'' \vert V_{r}\vert  \varepsilon'; i'  \rangle
   c_{\nu,\varepsilon' i'} (t)\,  t^{\nu', \varepsilon'' \varepsilon''_a i'' j''}_{\nu, \varepsilon' i'}  \,b_{\nu',\varepsilon'' \varepsilon''_a i'' j''}({\bf R}) .
\end{eqnarray*}
\end{widetext}
Note in the second line, that from the sum over all the vibrational states in the core-hole state ($\nu$) and the final dication state ($\nu'$), the energy conservation imposed by the delta function fixed the value of the Auger electron energy. Therefore, for every $(\nu,\nu')$ we have a different Auger electron energy determined by the energy conservation. If we assume that the Auger matrix transitions $\langle  \varepsilon'' \varepsilon''_a ; i'' j'' \vert V_{r}\vert  \varepsilon'; i'  \rangle$ are slowly dependent on Auger energy electron $\varepsilon_a$, and the Frank-Condon-factors related products $t^{\nu', \varepsilon'' \varepsilon''_a i'' j''}_{\nu, \varepsilon' i'}  \,b_{\nu',\varepsilon'' \varepsilon''_a i'' j''}({\bf R}) $ are also slowly dependent on $\varepsilon_a$ (as it is expected because they should be mainly dependent on the pes of the electronic core-hole and dication levels), we can finally derive the second equation of the EOMs (\ref{EOM_Auger_adb}) by defining

\begin{widetext}
\begin{eqnarray*}
{\Gamma_{\varepsilon i,\varepsilon' i'} ({\bf R}) \over 2} &=& \pi \sum_{\varepsilon'' \varepsilon'_a} \sum_{i'' j''} \langle  \varepsilon; i \vert V_{r}\vert  \varepsilon'' \varepsilon''_a ; i'' j''  \rangle \, 
\langle  \varepsilon'' \varepsilon''_a ; i'' j'' \vert V_{r}\vert  \varepsilon'; i'  \rangle  \;\; \delta(E_{\varepsilon'' \varepsilon''_a ; i'' j''} - E_{\varepsilon' ; i'}) \\
R_{\varepsilon i,\varepsilon' i'}({\bf R}) & = & -i \sum_{i'' j''} \langle  \varepsilon; i \vert V_{r}\vert  \varepsilon'' \varepsilon''_a ; i'' j''  \rangle \, 
\langle  \varepsilon'' \varepsilon''_a ; i'' j'' \vert V_{r}\vert  \varepsilon'; i'  \rangle \;\; P \left[ {1 \over E_{\varepsilon'' \varepsilon''_a ; i'' j''} - E_{\varepsilon' ; i'}}\right]
\end{eqnarray*}
\end{widetext}

Similary, we can use the adiabatic approximation in the ionization step in order to derive the first equation of the EOMs (\ref{EOM_Auger_adb}). First, we take the integral form of the second equation (\ref{EOM_Auger_adb}) and we substitute it into Eq. (\ref{EOM_Auger_RPA_1}). We obtain an integral over $t'$. Following a similar procedure than in the previous calculations, we divide the time-dependent factors in slow and fast time variant. Then we perform the integration and we obtain a ionization rate $\Gamma_I$ and a Stark shift $R_I$ factor.



\end{document}